\documentclass[pre,eqsecnum,showpacs]{revtex4}
\usepackage{amsfonts,amssymb,amsmath,bm}

\DeclareMathOperator{\Tr}{Tr}

\begin{document}
\title{Dynamics of inertial particles in a random flow with strong permanent shear}

\author{Grigory A. Sizov}

\affiliation{Landau Institute for Theoretical Physics RAS, \\
 119334, Kosygina 2, Moscow, Russia, \\
 and Moscow Institute of Physics and Technology.}

\date{\today}

\begin{abstract}
We consider advection of small inertial particles by a random fluid flow with a strong steady shear component. It is known that inertial particles suspended in a random flow can exhibit clusterization even if the flow is incompressible. We study this phenomenon through statistical characteristics of a separation vector between two particles. As usual in a random flow, moments of distance between particles grow exponentially. We calculate the rates of this growth using the saddle-point approximation in the path-integral formalism. We also calculate correction to the Lyapunov exponent due to small inertia by a perturbation theory expansion.
\end{abstract}


\maketitle
\section{Introduction}
 \label{sec:intro}

Let us consider a system of small identical inertial particles suspended in a random fluid flow. The flow consists of random fluctuations of a fluid velocity on the background of a strong permanent shear. This system can serve as a model of various natural and experimental situations: plankton colonies in the ocean \cite{90SF} or water droplets in warm clouds (which is useful for estimating rain initiation times \cite{02FFS,02FSV}). The most dramatic common feature of those situations is  clusterization of inertial particles advected by a random flow.

At first sight, one would expect an incompressible random flow to mix advected particles. However, due to inertia particles' velocity field differs from the fluid velocity field and becomes compressible. This compressibility leads to "unmixing"\cite{07WMOD}: emergence of clusters and voids in an initially uniform distribution of particles \cite{90SF,02FFS}.

Up to now, effect of inertia on a dynamics of advected particles was studied for isotropic fluctuating flows. For example, aggregation of inertial particles in a random compressible flow in one, two and three dimensions was considered in \cite{03WM}, \cite{04MW} and \cite{05MWDWL} respectively. It was shown that small inertia reduces mixing, that means it produces negative corrections to The Lyapunov exponents, whereas large inertia can turn the largest Lyapunov exponent to zero. This is interpreted as a phase transition from aggregating to non-aggregating phase. It should be mentioned that since the velocity field of particles deviates from the fluid velocity field in the presence of inertia, a sum of the Lyapunov exponents is generally non-zero. Moreover, this sum is always non-positive and can be interpreted as minus the rate of the mean entropy production per unit volume \cite{01FGV,97Rue}. The statistics of the concentration fluctuations was studied in \cite{01BFG}. In \cite{05Bec} the corresponding multifractal set in phase space was studied using concepts from dissipative dynamical systems theory (as opposed to behavoir of passive tracers, which can be thought of as a conservative dynamical system).

However, in many experimental and natural systems fluctuating flow does not exist alone, but against the background of a strong shear that changes slowly.
Processes we are interested in take place at scales much smaller than the viscous scale of turbulence, so the velocity field is spatially smooth.
We model it by a permanent shear flow with superposed weak gaussian spatially-uniform fluctuations of velocity gradients. Fluctuating part of the flow velocity is thereby short-correlated in time and linear in space.

Dynamics of tracers (particles without inertia) in a shear flow combined with weak fluctuations was studied in \cite{05CKLT,07Tur}.
The pure shear flow doesn't produce exponential growth of the distance between tracers, the distance grows only linearly. However, when combined with weak fluctuations of a flow velocity, the separation between the particles exhibits a non-trivial dynamics.

Consider two particles suspended in such a flow. Because of the shear component of the flow  one of them is moving on average faster than another (in this paragraph we project all the motion on the direction of the shear velocity). As the shear flow makes the first particle move faster, most of the time it is situated to the right of the second particle. However, from time to time noise can overtake the average velocity difference and place the second particle to the right of the first. This situation will persist for the short period of time, after which the first particle will again outrun the second. For separation vector this means the following behavior: most of the time it spends in the region of small angles of order $\phi_0=(D/s)^{1/3}$ along the shear direction, where $s$ is shear rate and $D$ is intensity of fluctuations. From time to time, with an average period $T\sim (Ds^2)^{-1/3}$ it tumbles, making a half-turn and again for a long time it stays directed along the shear velocity.

 This dynamics produces a stationary angular PDF that is mainly localized at small positive angles. However, due to tumbling,  it also has an algebraic tail ${\cal P}(\phi)\propto \phi^{-2}$ at $\phi_0\ll\phi\ll1$. The Lyapunov exponent is proportional to $(Ds^2)^{1/3}$ and $n$-th moment of distance grows exponentially with the rate $(n^4Ds^2)^{1/3}$ for $n\gg 1$
\cite{07Tur}.

The task of this paper is to study a system, in which both effects are present: particles have small inertia, and the two-dimensional random flow has a strong permanent shear component.
In our model there are four parameters that have dimension of time: the inverse fluctuations intensity $D^{-1}$, which is much large then the inverse shear rate $s^{-1}$, which is in turn much larger than the measure of particle inertia $\tau$. The smallest is the correlation time of fluctuations, which we completely ignore in this paper, treating fluctuations as a white noise.

 In Section \ref{sec:lap} we calculate a correction to the  Lyapunov exponent due to particles' inertia.
 We derive the system of stochastic equations, from which the Lyapunov exponent is expressed as an average of one of the variables.
 The corresponding Fokker-Planck equation is then transformed to a Schr\"odinger equation of a harmonic oscillator with a non-hermitian perturbation. We then perform perturbation theory calculations. The correction to the Lyapunov exponent turns out to be proportional to $-\tau \lambda^2$, where $\lambda\sim (Ds^2)^{1/3}$ is the Lyapunov exponent of the tracer.
 In Section \ref{sec:inst} we calculate the growth rates $\lambda_n$ of high moments of a separation between two particles. For large moment numbers this can be done analytically in a saddle-point approximation using path-integral formalism \cite{96FKLM}. We reproduce  the result from inertialess case $\lambda_n=(n^4 Ds^2)^{1/3}$ of \cite{07Tur} and find the correction to it due to a small inertia, which is $-2n \tau(nDs^2)^{2/3}$.
Finally, in Section \ref{sec:conclusion} we summarize the results and discuss possible paths of a further work.
\section{Lyapunov exponent}
\label{sec:lap}
The Lyapunov exponent is defined as the limit $\lambda =\lim\limits_{t\rightarrow\infty} t^{-1}\ln R(t)$.
It can be expressed from the rates of exponential growth of the distance,  $\lambda_n=\lim\limits_{t\rightarrow\infty} t^{-1}\ln \left\langle R^n\right\rangle $, at small $n$:
\begin{equation}
\lambda=\frac{d\lambda_n}{dn}|_{n=0}.
\label{smalln}
\end{equation}
For tracers this quantity was calculated in \cite{07Tur}. In this section we calculate the first correction to it due to a small inertia.

Let us describe here a formal model, in the framework of which we'll work further.
Velocity field, in which particles are suspended, consists of two components: permanent shear flow along $x$-axis
\begin{equation}
v_x=sy
\end{equation}
 and fluctuating part $v_i=\sigma_{ij} R_j$.
So the full tensor of the  velocity gradients is as follows:
\begin{equation}
 \Sigma_{ji}(t)=s\delta_{jx}\delta_{iy}+\sigma_{ji}(t).
 \label{sig}
 \end{equation}
It should satisfy the incompressibility condition $\Tr \hat \Sigma=0$.

Fluctuating part of the velocity is assumed delta-correlated in time, isotropic and incompressible
\begin{eqnarray}
 \langle \sigma_{ik}(t_1\!)\sigma_{jn}(t_2\!)\rangle
 \!=\!D(3\delta_{ij}\delta_{kn}\!-\!\delta_{ik}\delta_{jn}
 \!-\!\delta_{in}\delta_{jk})\delta(t_1\!-\!t_2), \quad
 \label{isotropic}
 \end{eqnarray}
    It is worthy of notice that in a shear flow the only relevant component of $\hat\sigma$ is $\sigma_{21}$, so the isotropic tensor structure of (\ref{isotropic}) is chosen only for convenience \cite{11LKS}.
The fluctuations are assumed weak relative to the shear, that is $D\ll s$.

Spherical particles in an incompressible flow $\mathbf{u}(\mathbf{r},t)$ are subject to Stock's drag force $\mathbf{f}=6\pi \eta  a(\mathbf{u}(\mathbf{r})-\mathbf{\dot r})$. Consider two such particles with a separation $\mathbf{R}(t)$ between them. Velocity field is smooth in our model, so in the first approximation $\mathbf{u}(\mathbf{r}+\mathbf{R})-\mathbf{u}(\mathbf{r})=\hat \Sigma \mathbf{R}$.
This leads to the following motion equation for $\mathbf{R}$
\begin{equation}
\tau\mathbf{\ddot R}+ \mathbf{\dot R}=\hat \Sigma \mathbf{R}.
\label{Req}
\end{equation}

Here $\tau=m/(6\pi \eta  a)$ is a measure of particle's inertia, which we assume small, namely $\tau\ll D$. It has a dimension of time and can be interpreted as the "response time" of a particle. We assume that all particles are identical and don't interact with each other.

 We pass to new variables: an angle $\phi$ respective to the shear direction, $\rho=\ln R$ and $Y_1=\dot\rho, Y_2=\dot \phi$. Equation (\ref{Req}) transforms to

\begin{equation}
\begin{cases}
\tau \dot Y_1=-Y_1+\tau(Y^2_2-Y_1^2)+s\sin \phi\cos\phi +\xi_1 \\
\tau \dot Y_2=-Y_2-2\tau Y_1 Y_2-s\sin^2\phi +\xi_2  \\
\dot \phi=Y_2
\end{cases}
\label{inertia_motion_eq}
\end{equation}
Though in (\ref{Req}) noise was multiplicative, in (\ref{inertia_motion_eq}) it has become additive: $\xi_1, \xi_2$ are independent white noises that are linear combinations of elements of $\hat\sigma$ (the price we pay for it is that the equations become nonlinear).
As long as the dynamics of $\phi,Y_1,Y_2$ is governed by the system of stochastic equations (\ref{inertia_motion_eq}), their joint PDF
satisfies Fokker-Planck equation $\tau \partial_t P(\phi,Y_1,Y_2,t)=\hat L P(\phi,Y_1,Y_2,t)$. The Fokker-Plank operator $\hat L$ is

\begin{equation}
\hat L=-\tau Y_2\partial_\phi +\partial_1\left(Y_1-s\sin \phi\cos\phi + \frac{D}{\tau}\partial_1\right)+\partial_2\left(Y_2+s\sin^2\phi+\frac{D}{\tau}\partial_2\right)+\tau \partial_1\left(Y_1^2-Y_2^2\right)+2\tau Y_1\partial_2 Y_2
\label{fokker_op}
\end{equation}
 $P(\phi,Y_1,Y_2)$ finally takes a stationary form, while $\rho$ grows infinitely with a stationary stochastic increment. Thus we look for zero modes of $\hat L$.

In order to make perturbation theory calculations more convenient, we pass from the Fokker-Planck to a Shr\"odinger equation using a substitution
\begin{equation}
P(\phi,x_1,x_2)=Q(\phi,x_1,x_2)\exp\left(-x_1^2/4-x_2^2/4\right),
\label{subst}
\end{equation}
where
$
x_1=\sqrt{\tau/D}\left( Y_1-s\sin \phi\cos\phi\right),  \  x_2=\sqrt{\tau/D}\left(Y_2+s\sin^2\phi\right)$.

Operator $\hat L$ transforms as follows:
\begin{center}
$\hat H\equiv- e^{(x_1^2+x_2^2)/4}\hat L e^{-(x_1^2+x_2^2)/4}=
-\epsilon\left[4x_1-x_2^3/2-x_1x_2^2/2-x_2\partial_\phi+2x_1 x_2\partial_2+(x_1^2-x_2^2)\partial_1+\right.$
$\left.+\alpha\left((4- x_1^2+2 x_1\partial_1)\frac{\sin 2\phi}{2}-\frac{x_1 x_2}{2}+\phi^2\partial_\phi-2\phi^2x_1\partial_2+x_2\partial_1+2\phi^2 x_2\partial_1\right)
\right]-
\left(\Delta - \frac{x_1^2+x_2^2}{4}+1\right)$,
\end{center}
where
 $\epsilon=\sqrt{D\tau},\  \alpha=s\sqrt{\tau/D}$. We assume that $\epsilon\ll 1,\  \epsilon\alpha\ll 1$.

The resulting hamiltonian $\hat H$ corresponds to a two-dimensinal harmonic oscillator with a non-hermitian perturbation. We express it through oscillator creation-annihilation operators $\hat a=x/2+\partial_x,\  \hat a^\dag=x/2-\partial_x$

\begin{equation}
\begin{cases}
\hat H^{(0)}=a_1^{\dag} a_1+a_2^\dag a_2 \\
\hat H^{(1)}=-\epsilon\left[-x_2\partial_\phi -a_1^\dag a_1^2-2 {a_1^\dag}^2 a_1-{a_1^\dag}^3+a_1^\dag (a_2^2-{a_2^\dag}^2)-2a_2^\dag a_1(a_2+a_2^\dag) \right]\propto \sqrt{\tau} \\
\hat H^{(2)}=-\epsilon\alpha\left[2\sin 2\phi-a_1^\dag a_2+2 a_1 a^\dag_2\sin^2\phi+\sin^2\phi\partial_\phi-a_1^\dag a_2^\dag\cos2\phi-x_1a_1^\dag\sin 2\phi\right] \propto \tau
\label{H}
\end{cases}
\end{equation}
Below in this section we will find corrections to a ground state wave function by perturbation theory expansion in small parameter $\epsilon$.
Let us notice that our perturbation theory is degenerate, because $\hat H^{(0)}$ acts as identity on the space of functions of $\phi$ and thus its eigenfunctions can have an arbitrary angular dependance.

 We look for a new ground state as a sum over non-perturbed oscillator eigenfunctions
 \begin{equation}
Q(x_1,x_2,\phi)=\sum\limits_n f_n(\phi) |n\rangle.
\label{series}
\end{equation}

 Further we denote by $f^{(k)}_{n_1n_2}$ a correction of $k^{th}$ order to $f(\phi)$ corresponding to the oscillator eigenstate with quantum numbers $n_1,n_2$. Obviously $f_{00}\sim 1$, and all other $f_i$ are small at least as $O(\epsilon)$. As we will see below, an equation that determines the proper zero-approximation wave function coincides with stationary Fokker-Planck equation for inertialess problem.

  We substitute series (\ref{series}) into $\hat H Q=0$ and obtain the following system of equations

\begin{equation}
\left(H_{nk}+E_k^{(0)} \delta_{nk}\right)f_k(\phi)=0, \  E_k^{(0)}=k_1+k_2.
\label{zerodet}
 \end{equation}

 Here we denote $H_{nk}\equiv\langle n_1 n_2 |\hat H|k_1 k_2\rangle$ - matrix elements of $\hat H$  between the non-perturbed oscillator states (they still act as operators on the space of functions of variable $\phi$).

We expand (\ref{zerodet}) successively in $\epsilon$ and obtain the following corrections to $f_n$ for $n_1+n_2>0$

\begin{equation}
\begin{cases}
f^{(1)}_n=- \frac{H_{n0}^{(1)}}{E_n} f^{(0)}_{00},
\\
f^{(2)}_n=\left(-\frac{H_{n0}^{(2)}}{E_n}+\sum \limits_k\frac{H_{nk}^{(1)} H_{k0}^{(1)}}{E_n E_k}\right)f_{00}^{(0)}
\\
f^{(3)}_n=\left(\sum \limits_k
\frac{H^{(2)}_{nk}H^{(1)}_{k0}+H^{(1)}_{nk}H^{(2)}_{k0}}{E_n E_k}-\sum \limits_{k,m}\frac{H^{(1)}_{nk}H^{(1)}_{km}H^{(1)}_{m0}}{E_n E_k E_m}\right)f_{00}^{(0)}- \frac{H^{(1)}_{n0}}{E_n}f^{(2)}_{00}
\label{third_ordercorr}
\end{cases}
\end{equation}
Here and in what follows we sums $n,k,m$  are over all oscillator states except the ground state.
Let us notice that $H^{(2)}$ preserves parity of sum of oscillator quantum numbers $n_1+n_2$, and $H^{(1)}$ inverts it, so $f_n$ with even $n_1+n_2$ have corrections only of even order in $\sqrt\tau$, and vice versa.

It's a bit more difficult to calculate $f_{00}$.
Expending (\ref{zerodet}) up to the 2nd order, we get an equation for $f_{00}^{(0)}$:
\begin{equation}
H_{00}^{(2)}f^{(0)}_{00}=\frac{H_{0k}^{(1)}H_{k0}^{(1)}}{E_k}f^{(0)}_{00}.
 \label{f000}
\end{equation}
Non-zero matrix elements entering this expression are $\langle 00|H|00\rangle=-\alpha\partial_\phi\sin^2\phi$ and $
\langle 00|H|01\rangle=\langle 01|H|00\rangle=-\epsilon \partial_\phi$.
Thus (\ref{f000}) coincides with inertialess Fokker-Plank equation for an angular PDF
\begin{equation}
\partial_\phi(s\sin^2\phi+D\partial_\phi)f_{00}^{(0)}=0
\label{inertia_less}
\end{equation}
This equation was analyzed in \cite{07Tur}.
In the region of small angles its normalized solution is $(s/D)^{1/3}F\left((s/D)^{1/3}\phi\right)$, where
\begin{equation}
F(\eta)=C\int \limits_0^\infty d\xi \exp\left(-\eta^3/3-(\xi-\eta)^3/3\right),\  C=\frac{2\sqrt{\pi}\Gamma(1/6)}{3^{5/6}}
\end{equation}
With this PDF, one can show that the inertialess Lyapunov exponent $\lambda=s\langle\phi\rangle$ is
\begin{equation}
 \lambda=\frac{\sqrt\pi\, 3^{1/3}}{\Gamma(1/6)} D^{1/3}s^{2/3}.
 \label{fok3}
\end{equation}
In order to calculate a correction to this exponent, we have to expand (\ref{zerodet}) to the 4th order.
Corrections of odd order to $f_{00}$ vanish, the first correction to $f_{00}$ is of the second order and satisfies the equation

\begin{equation}
H_{00}^{(2)}f^{(2)}_{00}=-\left(\sum\limits_{n,m}
 \frac{ H^{(1)}_{0n} H^{(2)}_{n m}  H^{(1)}_{m_0}+ H^{(1)}_{0 n} H^{(1)}_{n m}  H^{(2)}_{m_0}+ H^{(2)}_{0 n} H^{(1)}_{n m}  H^{(1)}_{m_0} }{E_n E_m}-\sum\limits_{n,m,k} \frac{H^{(1)}_{0n}H^{(1)}_{nm}H^{(1)}_{mk}H^{(1)}_{k0}}{E_n E_m E_k}\right) f^{(0)}_{00}
 \label{f002}
\end{equation}

Operator in the right-hand side is
\begin{equation}
-\tau^2\left(Ds\left(-\partial_\phi\sin 2\phi\partial_\phi+\partial_\phi\cos2\phi-\partial_\phi\sin\phi\partial_\phi\sin\phi\partial_\phi\right)- D^2\left(2\partial^2_\phi+\partial^4_\phi\right)\right)
\end{equation}
The term with second derivative renormalizes the fluctuations power $D$ : $D'=D(1+2 D\tau)$.

We expand (\ref{f002}) in small $\phi$ and search for a solution in the form $f^{(2)}_{00}=(s/D)^{1/3}F^{(2)}\left((s/D)^{1/3}\phi\right)$.
In the main order in $D/s$:

\begin{equation}
\left(\partial_\eta \eta^2+\partial_\eta^2\right)F^{(2)}(\eta)=-\tau(Ds^2)^{1/3}\left(-\partial_\eta+\partial_\eta\eta\partial_\eta\eta\partial_\eta+2\partial_\eta\eta\partial_\eta+\partial^4_\eta\right)F^{(0)}(\eta)
\label{eqF2}
\end{equation}

We look for a solution of (\ref{eqF2}) that doesn't violate normalization condition, that means $\int\limits_0^{2\pi} d\phi F^{(2)}(\phi)=0$. Its explicit form is
\begin{equation}
F^{(2)}(\eta)=\tau(Ds^2)^{1/3}e^{-\eta^3/3}\int\limits_{-\infty}^\eta d\eta_1 e^{\eta^3_1/3}
\int\limits_{-\infty}^{+\infty} d\eta_2 e^{-\eta^3_2/3}\int\limits_{-\infty}^{\eta_2} d\eta_3 e^{\eta^3_3/3}\left(h(\eta_1)-h(\eta_3)\right),
\end{equation}
where $h(\eta)=\left(1-3\eta\partial_\eta-\eta^2\partial_\eta-\partial^3_\eta\right)F^{(0)}(\eta)$

Our aim is the Lyapunov exponent, which is given by
\begin{equation}
\lambda=\langle Y_1\rangle=\sqrt{D/\tau}\langle x_1 \rangle+s\langle\phi\rangle
\label{Y1}
\end{equation}
One can see that the correction to The Lyapunov exponent comes from two sources: correction to the mean angle and non-zero $\langle x_1\rangle$.
The correction to the mean angle  can be calculated from the correction to the angular PDF
\begin{equation}
\langle \phi \rangle=\int dx_1 dx_2 d\phi \phi \exp\left(-\frac{x_1^2+x_2^2}{4}\right)Q(x_1,x_2,\phi)=\int\limits_0^{2\pi} \phi f_{00}(\phi)d\phi=\langle\phi\rangle_0+\int \limits_0^{2\pi} \phi f^{(2)}_{00}(\phi)d\phi
\end{equation}

$\delta \phi=(D/s)^{1/3}\int d\eta \eta F^{(2)}(\eta) =-C_1 \tau(D^2 s)^{1/3}$.

\begin{equation}
C_1=-\int \limits_{-\infty}^{+\infty}d\eta \eta e^{-\eta^3/3}\int\limits_{-\infty}^\eta d\eta_1 e^{\eta^3_1/3}
\int\limits_{-\infty}^{+\infty} d\eta_2 e^{-\eta^3_2/3}\int\limits_{-\infty}^{\eta_2} d\eta_3 e^{\eta^3_3/3}\left(h(\eta_1)-h(\eta_3)\right)\approx 9.8
\end{equation}
So, increment correction due to change of the mean angle is $\delta\lambda_1=s\delta\phi=-C_1\tau (Ds^2)^{2/3} $

The second increment correction is related to the fact that averaged $Y_1$ is no longer equal to $s\langle \phi\rangle$ since we take inertia into account

\begin{equation}
\langle x_1 \rangle=\int dx_1 dx_2 d\phi x_1 \exp\left(-\frac{x_1^2+x_2^2}{4}\right)Q(x_1,x_2,\phi)=\int\limits_0^{2\pi} f_{10}(\phi) d\phi
\end{equation}
Full derivatives vanish after an integration, so the first order correction $f^{(1)}_{10}\sim \partial_\phi f^{(0)}_{00}$ doesn't contribute to the answer. The second-order correction to $f_{10}$ is zero, consider the third order (\ref{third_ordercorr}):
\begin{equation}
f^{(3)}_{10}=-\left(\epsilon^2\alpha\partial_\phi\left(1+\frac{1}{2}\cos 2\phi\right)+\epsilon^2\alpha\sin 2\phi+\frac{2}{3}\epsilon^3\partial_\phi^2\right)f_{00}^{(0)}
\end{equation}
\begin{equation}
\langle x_1 \rangle=-\epsilon^2\alpha\int\limits_0^{2\pi} d\phi \sin 2\phi f_{00}^{(0)}(\phi)
\end{equation}
We see that correction to the Lyapunov exponent due to non-zero $\langle x_1\rangle$ is much smaller than the correction due to change of the mean angle and can be neglected as long as $D/s\ll 1$:
\begin{equation}
\delta\lambda_2=\sqrt{\frac{D}{\tau}}\langle x_1\rangle=-2Ds\tau \langle \phi\rangle=-2D\tau (Ds^2)^{1/3}\ll \delta\lambda_1.
\end{equation}

Thus we have calculated the correction to the first Lyapunov exponent. The correction is negative, so inertia reduces mixing like in an isotropic situation.
Using result for inertialess Lyapunov exponent $\lambda=\frac{\sqrt\pi\, 3^{1/3}}{\Gamma(1/6)} D^{1/3}s^{2/3}$ from \cite{07Tur} we can represent the answer in the relative form
\begin{equation}
\delta \lambda/\lambda=-\frac{C_1\Gamma^2(1/6)}{\pi 3^{2/3}}\lambda\tau\approx -46.5\lambda\tau
\end{equation}

\section{Large n saddle-point approximation}
\label{sec:inst}
In the previous section we have calculated the Lyapunov exponent, which is determined by the growth rates $\lambda_n$ of $\langle R^{n}(t)\rangle$ at small $n$ (\ref{smalln}), and in this section we'll calculate growth rates for large $n$.
 The method we exploit to do this is based on the Optimal Fluctuation principle (saddle-point approximation in a functional space).
  Its application to problems of statistical hydrodynamics was developed in papers \cite{96FKLM,99VL, 98BL}. To make the saddle-point approximation more transparent, we use path-integral formalism for averaging over trajectories of stochastic processes.

First we develop a path-integral representation of $R^{2n}$ averaged over the velocity statistics. Random velocity enters equations (\ref{inertia_motion_eq}) through two independent white-noise stochastic processes, so we can express the average as
\begin{equation}
\langle R^{2n}(t)\rangle=\int  D\xi_1D\xi_2 R^{2n}{\cal P}[\xi_1]{\cal P}[\xi_1]\propto e^{\lambda_{2n} t}
\label{fint}
\end{equation}
where ${\cal P}$ is a probability measure of a white noise
\begin{equation}
{\cal P}[\xi]=\exp\left(-\frac{1}{2D}\int\limits^t_0 dt'\xi^2\right).
\end{equation}
We express the noise from Langevin equations (\ref{inertia_motion_eq}) and the functional integral over the noise $(\ref{fint})$ transforms to the functional integral over $\phi, Y_1$

\begin{equation}
\left\langle R^{2n}(t)\right\rangle=\int  DY_1  D\phi  \exp\left(-S[Y_1,\phi]\right)
\label{fint_S}
\end{equation}

with an action determined by the original Langevin equations

\begin{equation}
S[Y_1,\phi]=\int\limits_0^t  dt' \left[\left(-\tau \dot Y_1-Y_1+\tau(\dot\phi^2-Y_1^2)+s\sin \phi\cos\phi\right)^2+\left(-\tau \ddot \phi-\dot\phi-2\tau Y_1 \dot\phi-s\sin^2\phi\right)^2+2nY_1\right].
\label{action}
\end{equation}

At large times and $n\gg1$ the main contribution to this integral is given by the saddle-point approximation \cite{96FKLM}. This means that we calculate optimal
trajectory $\phi^\star(t),Y_1^\star(t), Y_2^\star(t)$ which minimizes $S$, then in the main approximation (\ref{fint_S}) equals to $\exp\left(-S[\phi^\star,Y_1^\star, Y_2^\star]\right)$.
At first we will follow this procedure for slightly more general form of Langevin equations than (\ref{inertia_motion_eq}) in order to use it for both inertial and inertialess cases.

Consider the following system of stochastic equations
\begin{equation}
\begin{cases}
Y_1=f_1(\phi)+\xi_1 \\
\dot \phi=f_2(\phi)+\xi_2,
\end{cases}
\end{equation}
where $\xi_1,\xi_2$ are independent white noises.
The corresponding action is
\begin{equation}
S\left[\phi(t),Y_1(t)\right]=\frac{1}{2D}\int\limits_0^t dt'\left(\xi_1^2+\xi_2^2\right)-2n\rho=\int \limits_0^t dt'\left(\frac{1}{2D}\left(Y_1-f_1(\phi)\right)^2+\frac{1}{2D}\left(\dot\phi-f_2(\phi)\right)^2-2nY_1\right)
\end{equation}
Motion equations and boundary conditions are obtained by varying this action with respect to $\delta Y_1, \delta\phi$
\begin{equation}
\begin{cases}
Y_1-f_1=2nD, \\
\ddot\phi=f_2\frac{\partial f_2}{\partial \phi}-2nD \frac{\partial f_1}{\partial\phi},\\
\left(\dot\phi-f_2(\phi)\right)|_{0,t}=0.
\end{cases}
\end{equation}
The second equation describes Newtonian motion in the potential
\begin{equation}
U(\phi)=2nDf_1(\phi)-f_2^2(\phi)/2,
\label{potential}
\end{equation}
 thus we can introduce conserved "energy"
\begin{equation}
(\dot\phi)^2/2+U(\phi)=E.
\end{equation}

Now we have solution for optimal trajectory in an implicit form $t=\int \frac{d\phi}{\sqrt{E-U(\phi)}}$ and substitute it into the action
\begin{center}
$S=-\int\limits_0^t dt'\left(-\frac{1}{2D}\left(f_2(\phi)-\sqrt{2E-4nDf_1(\phi)+f_2^2(\phi)}\right)^2+2Dn^2 +2nf_1\right)=$
$=\frac{1}{D}\int \limits_0^t dt'\left(f_2^2(\phi)-4nDf_1(\phi)\right)-2Dn^2t-\frac{1}{D}f_2(\phi)\phi|_0^t+Et/D$
\end{center}
Using energy conservation relation, the first term can be rewritten as $2U=2E-\dot \phi^2$.
Finally the growth rates are expressed as follows
\begin{equation}
\lambda_{2n}t=-\frac{1}{D}\int\limits_0^t dt'\dot\phi^2+\frac{1}{D}f_2(\phi)\phi|_0^t+Et/D+2Dn^2t
\label{action_final}
\end{equation}
Below we apply this formula to concrete $f_1, f_2$ corresponding to inertialess and inertial dynamics.

 In the inertialess case ($\tau=0$) from (\ref{inertia_motion_eq}) one can see that $f_1(\phi)=s\sin\phi\cos\phi,f_2(\phi)=-s\sin^2\phi$. The potential $U(\phi)=nDs\sin2\phi-s^2\sin^4\phi/2$ has a maximum at $\phi_m=(nD/s)^{1/3}$.  Boundary conditions require that
 \begin{equation}
 \sin {2\phi(0)}=\sin {2\phi(t)}=\frac{E}{nDs},
 \label{boundary}
 \end{equation}
  so the optimal trajectory is tumbling from some initial positive $\phi(0)$ to some angle $\phi(t)=k\pi+\phi(0)$. As long as we keep time $t$ is fixed, action doesn't depend on the number of turns $k$.
 We investigate behavior of the system at large times, so optimal trajectories which we consider should take long time. This is achieved if a gap between the energy and the maximum of the potential is small
: $t\sim \ln(E-U(\phi_m))$, so $E\approx U_{max}=\frac{3}{2}(n^4D^4s^2)^{1/3}$.

  The first term in the action (\ref{action_final}) doesn't grow with time, because it can be estimated from above: $\int d\phi \sqrt{2(E-U(\phi))}<\pi k \cdot2\max\Delta U(\phi)$. So we neglect it as compared with $\lambda t$.
 The second term in (\ref{action_final}) vanishes due to boundary conditions (\ref{boundary}).
Finally, increment is determined by the energy alone
\begin{equation}
\lambda_{2n}=E/D=\frac{3}{2}(n^4 Ds^2)^{1/3}+2n^2D.
\label{incr_inertialess}
\end{equation}

Consider now a correction linear in $\tau$. Formally, order of the motion equation increases, but all terms with high derivatives are multiplied by small parameter, so we can substitute all higher derivatives from the inertialess solution
\begin{equation}
\dot\phi=\sqrt{2(E-U_0(\phi))},\   \ddot\phi=-U'_0(\phi)
\label{deriv}
\end{equation} and prevent the order of the motion equation from increasing.
Increment is still determined by an energy, which is very close to maximum of the potential.
\begin{equation}
U(\phi_m)=U_0(\phi_m^{(0)}+\delta\phi)+\delta U(\phi_m^{(0)}+\delta\phi)=U_0(\phi_m)+\delta U(\phi_m^{(0)})+O(\tau^2)
\end{equation}

From (\ref{potential}) and (\ref{inertia_motion_eq}) follows
 \begin{eqnarray}
\nonumber
\delta U=2nD\delta f_1-f_2\delta f_2, \\
\nonumber
\delta f_1=-\tau(Y_1^2+\dot Y_1-\dot\phi^2), \\
\delta f_2=-\tau (2 Y_1 \dot \phi+\ddot \phi).
\label{df}
\end{eqnarray}
Here all derivatives of $\phi(t)$ are assumed to be expressed through $\phi(t)$ using (\ref{deriv}).

Let us simplify (\ref{df}) when $\phi=\phi_{max}$. At this point $\ddot\phi=0$, because $\phi(t)$ obeys an equation  $\ddot\phi=-U'(\phi)$. Since the gap between energy and potential maximum is small and $\dot \phi=\sqrt{E-U(\phi)}$, so $\dot\phi$ is also small.

 Substituting $\phi_{max}=(nD/s)^{1/3}$ into (\ref{Y1}), we obtain the value of $Y_1$ at $\phi_{max}$: $  Y_1=2nD+s(nD/s)^{1/3}.$
Again, the first term is generated by the shear flow and at $n\ll s/D$ dominates over the second, which is determined by fluctuations
   Using (\ref{Y1}) and the fact that $\dot\phi(\phi_m)=0$ we conclude that $\dot Y_1=0$.

   Now we collect it all together:

   $\delta f_2(\phi_m)=0$,

   $\delta f_1(\phi_m)=-\tau (2nD+s(nD/s)^{1/3})^2$.

Thus, we've calculated the correction to the potential maximum. It ultimately determines correction to growth rate of $\left\langle R^n\right\rangle$.
\begin{equation}
\delta \lambda_n=\delta E/D=2n \tau (2nD+s(nD/s)^{1/3})^2\approx -2n \tau(nDs^2)^{2/3}
\label{incr_correction}
\end{equation}

Let us represent the answer in the relative form
\begin{equation}
\delta\lambda_n/\lambda_n=-\frac{4}{3}\tau(n Ds^2)^{1/3}=-\frac{8}{9n}\tau \lambda_n.
\end{equation}

   Let us notice that correction (\ref{incr_correction}) depends on $n$ stronger than the main answer (\ref{incr_inertialess}). However, for $n\ll s/D$ it remains much lesser the main answer, because we assume $s\tau\ll 1$.
\section{Conclusion}
\label{sec:conclusion}
We have studied the dynamics of inertial particles in a two-dimensional random flow with a strong shear component.
We extracted information about the system from the statistical characteristics of the separation vector between two particles $\mathbf{R}(t)$. The moments of the distance grow exponentially in time and the growth rates depend non-linearly on $n$.
 We have calculated the first correction to the Lyapunov exponent, which is negative and of order $\tau\lambda^2$, where $\lambda$ is the inertialess Lyapunov exponent. This means that inertia reduces mixing and the relative correction to $\lambda$ is of order $\lambda\tau$.
  Let us note that for the shear-less situation, where inertialess Lyapunov exponent is $2D$ and the first correction due to inertia is $D\tau$, relative correction is also of order $\lambda\tau$ \cite{04MW}.

 We have also calculated $\lambda_n$ - the growth rates of the high moments of the inter-particle distance using a saddle-point approximation in the functional space (instanton method). The correction to the growth rate of the n-th moment due to inertia is found to be of order $-n \tau(nDs^2)^{2/3}$. Let us notice that the relative correction is proportional to $\delta\lambda_n/\lambda_n\sim n^{-1}\tau\lambda_n$, so the role of inertia reduces with $n$.

Let us mention a few issues that are interesting to investigate further.

As it is was shown in \cite{04MW}, in the isotropic model, as particles grow heavier, a phase transition to non-clustering phase may occur. We don't know yet if this transition exists in the presence of a strong shear.

The velocity fluctuations we considered were white-correlated in time. Obviously this is an approximation, because any real noise is colored. It's unclear if a finite correlation time of the noise can lead to new qualitative effects.

The Lyapunov exponents can be used to determine fractal characteristics of a cluster that is formed by particles, namely the so-called Lyapunov dimension \cite{10WMG,79KY}. For this purpose one has to  calculate corrections to both Lyapunov exponents (in two dimensions) up to the second order in $\tau$, because the first order cancels in fractal dimension.

In this paper we assumed that the parameters entering the model satisfy $D\tau\ll s\tau \ll 1$. Another limiting case, interesting for applications in atmospheric turbulence (though more complicated technically), is the case of larger inertia $D\tau\ll 1\ll s\tau$.

\acknowledgments
Author thanks G. Falkovich for formulating the problem, for numerous discussions and remarks on the text, V.V. Lebedev, I.V. Kolokolov and P.E Vorobev for helpful discussions from which this work undoubtedly benefited.

This research was supported by Dynasty Foundation Scholarship, Federal Targeted Program of RF "S\&S-PPIR" and Federal Targeted Program of RF "Kadry".

\end{document}